\documentclass{an}
\usepackage{graphicx}
\usepackage{times}
\usepackage{fancyhdr}
\usepackage{natbib}
\begin{document}

\title{New brown dwarf companions to dM(e) stars\thanks{This paper is partly based on observations obtained at the European Southern Observatory at La Silla, Chile under program ID 072.C-0571(B) and on observations collected at the Centro Astron{\'{o}}mico Hispano Alem{\'{a}}n (CAHA) at Calar Alto, operated jointly by the Max-Planck Institut f\"ur Astronomie and the Instituto de Astrof{\'{\i}}sica de Andaluc{\'{\i}}a (CSIC).}}

\author{Andreas Seifahrt\inst{1,2}
\and Markus Mugrauer\inst{2}
\and Manuela Wiese\inst{2}
\and Ralph Neuh\"auser\inst{2}
\and Eike W. Guenther\inst{3}}
\institute{
European Southern Observatory, Karl-Schwarzschild-Str. 2, 85748 Garching, Germany
\and 
Astrophysikalisches Institut und Universit\"ats-Sternwarte Jena, Schillerg\"asschen 2-3, 07745 Jena, Germany
\and 
Th\"uringer Landessternwarte Tautenburg, Sternwarte 5, 07778 Tautenburg, Germany}

\date{Received $<$date$>$; 
accepted $<$date$>$;
published online $<$date$>$}

\abstract{
We present new astrometric and spectroscopic data to confirm two new M/L dwarf systems, G124-62 and LHS5166, and discuss the nature of a third system (LP261-75). Age and thus mass determinations of the L dwarf companions are discussed based on various activity-age relationships of the M dwarf primaries. This publication will update the list of widely separated substellar companions to nearby stars.   
\keywords{stars: late-type --- stars: low-mass, brown dwarfs --- astrometry ---  binaries: general}
}

\correspondence{aseifahr@eso.org}

\maketitle

\section{Introduction}
Among the 460 brown dwarfs known today (see the M, L, and T dwarf compendium housed at DwarfArchives.org and maintained by Chris Gelino, Davy Kirkpatrick, and Adam Burgasser) only a minor fraction are companions to main sequence stars. These objects are of special interest since the stellar primary provides crucial information for the system like metalicity, distance and age, assuming coevality. With this information on hand one can constrain the mass of the companion that can not be determined out of direct observations, because the substellar nature of these objects causes a continuous decrease in temperature
and luminosity over time and thus a degeneracy in the determination of mass from models.

Moreover the fraction of binaries among these companions may give an insight into the question of the origin and evolution of brown dwarfs. Unfortunately only a small number of substellar companions is known today. Recently, \citet{Burgasser05} have presented a compilation of all known widely separated ($\geq$ 100 AU) stellar -- brown dwarf multiple systems. 

We intend to update this list with two new M/L dwarf systems and discuss the nature of a system already listed in Burgasser et al. (2005). 
\section{Data and analysis}
\subsection{G124-62 \& DENIS-P J1441-0945}
In \citet{Seifahrt05} we present new astrometric and photometric measurements of the high proper motion star G124-62 and the binary L1 dwarf DENIS-P J1441-0945 \citep{Martin99,Bouy03}. Analysis of archival HST images and spectro-photometric parallax measurements of DENIS-P J1441-0945 give a distance of $34\pm7$ pc and a proper motion close to that of G124-62. Comparison of SuperCOSMOS and 2MASS images confirms that these objects form a common proper motion pair.

High-resolution optical spectroscopy of G124-62 reveals a spectral type of dM4.5e. The kinematics of the system show that it is a member of the Hyades supercluster. The resulting age constraints for G124-62 are 500 -- 800 Myr, based on the reported absence of Li in the spectrum of G124-62 B, the x-ray luminosity of the primary and the (V-Ic) -- age relationship proposed by \citet{Gizis02}. Finally we derive the mass of each L dwarf component of DENIS-P J1441-0945 to be $0.072^{+0.010}_{-0.018}$~M$_{\sun}$, hence close to the hydrogen burning mass limit.

\subsection{LHS 5166 \& 2MASSW J1004-3335}
\begin{figure}[h!]
\resizebox{\hsize}{!}
{\includegraphics[bb=0 0 332 300,clip]{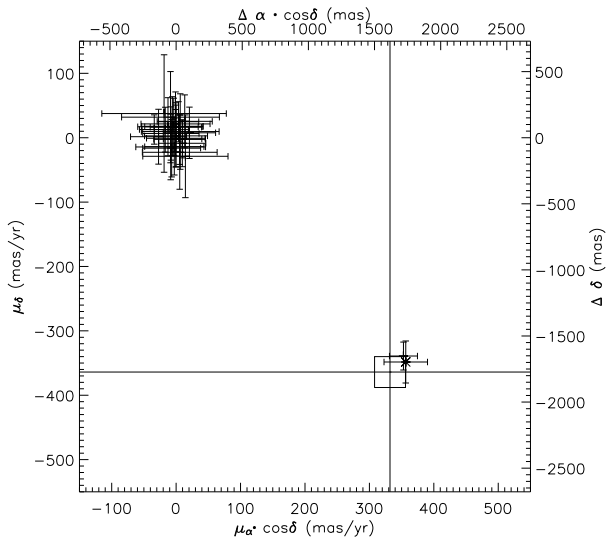}}
\caption{Measured position differences between 2MASS (1999/05/02) and SOFI (2004/03/15). 
LHS 5166 marked as a cross and 2MASSW J1004-3335 as a triangle. All background objects in the frame remain at zero proper motion. The expected proper motion for LHS 5166 from the NLTT catalogue is shown by a box. 1$\sigma$~errors are plotted for all values.}
\label{pm}
\end{figure}

\citet{Gizis02a} discovered the L4 dwarf 2MASSW J1004-3335 in the line of sight towards the TW Hya association. A companionship to the nearby high proper motion star LHS 5166 was proposed but not confirmed. 
We observed the field in March 2004 with SOFI at the NTT on La Silla and did follow-up spectroscopy of the companion (JHK, SOFI) and of the primary (VIS, EMMI). The astrometry clearly shows that both objects form a common proper motion pair (see Fig.~\ref{pm}). 

\begin{figure}
\resizebox{\hsize}{!}
{\includegraphics[bb=0 0 396 292,clip]{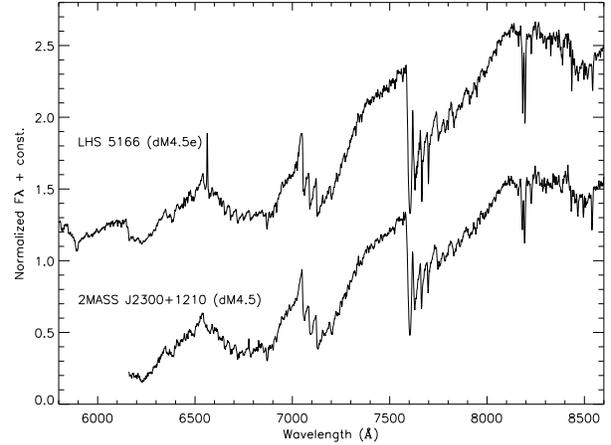}}
\caption{Low resolution (R$\sim$1100) optical spectra of LHS 5166 (upper spectrum) and a template dM4.5 (lower spectrum) taken with Keck/LRIS by \citet{Kirk99}. The H$\alpha$ emission peak of LHS 5166 is clearly visible.}
\label{spec}
\end{figure}

The low-res EMMI spectrum of LHS 5166 (R$\sim$1100) shows a clear H$\alpha$ emission feature with a equivalent width of $\sim$3.5 \AA\, (see Fig.~\ref{spec}). From TiO5 and VO-a \citep{Cruz02} as well as from PC3 \citep{Martin99} spectral indices we determine a spectral type of dM4.5e ($\pm$ 0.25).

After a careful revision of the published photometric and spectro-photometric distance relationships by \citet{Henry94,Kirk94,Cruz02,Scholz05,West05} and \citet{Henry04} we decided to use the latter publication as a reference to derive a photometric distance of LHS 5166 since it is based on the highest number of reference objects with measured parallax. From 
12 photometric distance relations given in \citet{Henry04} we derived a weighted mean of the distance of LHS 5166 A to $d=19.0\pm3.0$\,pc. 

Equivalently, we have calculated the spectro-photometric distance of LHS 5166 B (2MASSW J1004-3335) from relationships given by \citet{Dahn02,Cruz03,Kirk00} and \citet{Vrba04} to $d=19.0\pm2.0$\,pc, a perfect match. Hence LHS 5166 and 2MASSW J1004-3335 share the same proper motion and have the same distance within the 1$\sigma$ uncertainties.

To derive the mass of the L4 companion and probe its substellar nature we need the age of this object. This quantity can only be obtained from the primary, given coevality for both objects. For the lower limit we use the decay of x-ray activity of low-mass stars over time. The x-ray luminosity of LHS 5166 A, derived from the ROSAT Faint Source Catalog to be $\log L_x=27.91$, is lower than for all objects in the Hyades \citep{stelzer} even though LHS 5166 A is chromospherically active. This leads us to the conclusion that the object must be older than the Hyades cluster (625 Myr). An upper age limit of 2.6 Gyr can be estimated using the (V-Ic) -- age relationship by \citet{Gizis02} and the photometry from SPM and DENIS catalogue. 

\begin{figure}
\resizebox{\hsize}{!}
{\includegraphics[bb=0 0 332 249,clip]{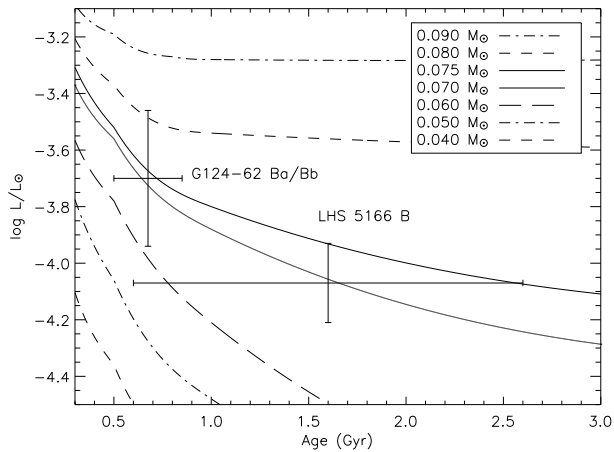}}
\caption{Luminosity as a function of age for various low mass stars and brown dwarfs from DUSTY models (Chabrier et al. 2000). G124-62Ba/Bb and LHS5166 B are shown with their 1$\sigma$?errors.}
\label{mass}
\end{figure}

Equipped with these quantities we derive the mass of LHS 5166 B (2MASSW J1004-3335) to $0.070^{+0.005}_{-0.015}$~M$_{\sun}$, thus like G124-62 Ba and Bb near the hydrogen burning mass limit (see Fig.~\ref{mass}). 

\subsection{LP 261-75 \& 2MASS J0951+3558}
As already mentioned by \citet{Burgasser05} 2MASS J0951+3558, a brown dwarf of spectral type L6 \citep{Kirk00} shares the proper motion of the nearby high proper motion star LP261-75.

2MASS J0951+3558 has a measured parallax of $\pi=16.09\pm7.4$ mas, hence a distance of $d=59^{+56}_{-17}$ pc \citep{Vrba04}. This value is only in marginally agreement with the mean spectro-photometric distance of $d=43\pm5$ pc derived from relationships for L6 dwarfs given by \citet{Kirk00,Dahn02,Hawley02,Cruz03,Vrba04} and \citet{Scholz05}.

We have obtained a high resolution optical spectrum of LP 261-75 (although of low S/N) with the FOCES spectrograph on Calar Alto and used the SDSS database (DR3) to determine the spectral type of LP 261-75 to dM4.0e ($\pm$0.5). The spectrum clearly shows emission features of H$\alpha$ and NaI, thus this object is chromospherically active (H$\alpha$ EW$\sim$2.75\,\AA\,). 

Although the kink of the main sequence at spectral type M4 prohibits a very precise spectro-photometric parallax measurement, the distance of LP 261-75 can by constrained to $d=28\pm5$ pc, using the calibrations of \citet{Hawley02,West05} and \citet{Scholz05}. Even with this rather conservative distance determination the difference in distance to the L6 dwarf is more than 2$\sigma$, leaving a significant doubt that both objects are physically bound. However, the likelihood of finding a brown dwarf and a M4.0 star separated only 13.3$\arcsec$ and having the same proper motion but being unbound (i.e. physically unconnected) is nearly negligible. More data on both objects have to be taken to test whether this system is real. 

Meanwhile we can determine age criteria for both objects. LP 216-75 is associated with a faint ROSAT source. We have reanalyzed the original ROSAT data to confirm the identity of the source since the original entry in the ROSAT Faint Source Catalog had an offset to the optical position of nearly 30$\arcsec$. Using the method described in \citet{Neu95} we determined a countrate of $0.037\pm0.011$ cts/s with a hardness ratio of $HR1=0.0\pm0.30$.
At a distance of $28\pm5$ pc this countrate translates into a flux of $\log Lx=28.46^{+0.33}_{-0.41}$. This flux is a typical value for the active M dwarfs in Hyades cluster \citep{stelzer}, hence for similar objects with an age of 625 Myr. From the (V-Ic) color of 2.9 we can additionally conclude on an upper age limit of about 2.2 Gyr for LP 216-75. 

Since \citet{Kirk00} found no LiI absorption in 2MASS J0951+3558 it can be concluded that the L6 dwarf is older than 0.5 Gyr. To combine these age limits and conclude on the mass of 2MASS J0951+3558 one has to prove that both objects are physically bound, a task that still needs to be done.

\section{Summary}
We have demonstrated in \citet{Seifahrt05} and in this publication that G124-62 and DENIS-P J1441-0945 as well as LHS 5166 and 2MASSW J1004-3335 form common proper motion pairs and are physically bound. Their L dwarf components have masses of $0.072^{+0.010}_{-0.018}$~M$_{\sun}$ and $0.070^{+0.005}_{-0.015}$~M$_{\sun}$, respectively.
Furthermore we have presented new data on LP 261-75 and 2MASS J0951+3558, a dM4.0e and a L6 dwarf with common proper motion. The spectro-photometric distance determination of both objects leaves doubt on their physical companionship making it necessary to collect further data on this pair to reveal its true nature.  

\acknowledgements
We would like to thank Laurent Cambresy, Observatoire de Strasbourg, for the preliminary photometric data on LHS 5166 from the Deep Near Infrared Survey of the Southern Sky (DENIS). This publication makes use of data products from the Two Micron All Sky Survey %(which is a joint project of the University of Massachusetts and the Infrared Processing and Analysis Center/California Institute of Technology, funded by the National Aeronautics and Space Administration and the National Science Foundation) 
and data based on the SuperCOSMOS Sky Surveys at the Wide Field Astronomy Unit of the Institute of Astronomy, University of Edinburgh. We have also used the VizieR catalogue access tool, CDS, Strasbourg and the SDSS data archive. Funding for the creation and distribution of the SDSS Archive has been provided by the Alfred P. Sloan Foundation, the Participating Institutions, the National Aeronautics and Space Administration, the National Science Foundation, the U.S. Department of Energy, the Japanese Monbukagakusho, and the Max Planck Society.

\end{document}